\newcommand{\bQ}{\mathbb{Q}}
\newcommand{\bR}{\mathbb{R}}
\newcommand{\bZ}{\mathbb{Z}}
\newcommand{\boxtt}[1]{\mbox{\texttt{#1}}}
\renewcommand{\epsilon}{\varepsilon}
\renewcommand{\phi}{\varphi}
\newcommand{\prof}{\noindent\(\clubsuit\)\ }
\newcommand{\qed}{\hspace*{3pt}\hfill\(\clubsuit\)}

\newcommand{\recordends}[2]%
{\path #1 -- coordinate [pos = 0] (c1) coordinate [pos = 1] (c2) #2 ;}

\newenvironment{fenum}{%
\vspace{-2\parskip}\begin{enumerate} \itemsep = -0.2em%
}{%
\end{enumerate}
}
\newenvironment{fitem}{%
\vspace{-2\parskip}\begin{itemize} \itemsep = -0.2em%
}{%
\end{itemize}
}

\newtheorem{thr}{Theorem}

\documentclass[a4paper,12pt]{article}
\usepackage{amsfonts}
\usepackage{amssymb}
\usepackage{amsmath}
\usepackage{calrsfs}
\usepackage{verbatim}
\usepackage{graphicx}
\usepackage[usenames,dvipsnames]{color}
\usepackage{url}
\usepackage{tikz}

 \DeclareMathOperator{\im}{im}

\definecolor{green}{rgb}{0,0.75,0}
\definecolor{violet}{rgb}{0.8225,0.2855,0.5225}
\definecolor{pink}{rgb}{1,0.5,0.5}
\definecolor{orange}{rgb}{0.8,0.6,0.3}

\title{About the Kannan--Bachem algorithm}
\author{Francis Sergeraert}
\date{\footnotesize\emph{October 2024}}

\begin{document}
\voffset=-2.5cm \hoffset=-0.9cm \sloppy


\maketitle


\flushright{\footnotesize\textit{The devil is in the details}}

\abstract{The Smith reduction is a basic tool when analyzing integer matrices  
up to equivalence, and the Kannan-Bachem (KB) algorithm is the first 
polynomial algorithm computing such a reduction. Using this algorithm in 
complicated situations where the rank of the studied matrix is not maximal 
revealed an unexpected obstacle in the algorithm. This difficulty is 
described, analyzed, a simple solution is given to overcome it, finally 
leading to a general organization of the KB algorithm, simpler than the 
original one, efficient and having a general scope.

An equivalent algorithm is used by the Magma program, without any detailed 
explanation, without any reference. The present text could so be useful.}

\section{Introduction.}

The Smith reduction of an \(n \times m\) integer matrix \(d: \bZ^m \rightarrow 
\bZ^n\) is a \emph{diagonal} matrix \(s: \bZ^m \rightarrow \bZ^n\) satisfying 
the following conditions:
\begin{fitem}
\item
The matrices \(s\) and \(d\) are \emph{equivalent}, that is, there exist two 
\emph{invertible} matrices \(u: \bZ^m \rightarrow \bZ^m\) and \(v: \bZ^n 
\rightarrow \bZ^n \) with \(s = vdu\).
\item
The non-null entries \(d_1, \ldots, d_k\) of the \(s\)-diagonal are positive 
and satisfy the \emph{divisor condition}: every entry \(d_i\) divides the next 
one \(d_{i+1}\); in particular the last one \(d_k\) divides the possible 0 in 
position \(k+1\) on the diagonal if the rank \(k\) is less than \(m\) and 
\(n\).
\end{fitem}%

The Smith reduction of integer matrices is used in many domains, in particular 
intensively used in computational Algebraic Topology. Calculating a homology 
group often consists in determining two boundary integer matrices \(d\) 
and~\(d'\) satisfying \(d'd = 0\) and the looked-for homology group is the 
quotient \(\ker d' / \im d\). Then the Smith reductions of the matrices \(d\) 
and~\(d'\) directly give the corresponding homology group.

In Constructive Algebraic Topology~\cite{rbsr06}, sophisticated computations 
often needing several weeks or months of runtime on powerful computers finally 
produce such matrices \(d\) and \(d'\), the corresponding homology group 
\(\ker d' / \im d\) being some homology or homotopy group unreachable by other 
means. 

Recently, such a calculation produced an integer matrix \(T_8\) of size \(684 
\times 1995\) and a matrix \(T_9\) of size \(1995 \times 5796\) with \(T_8 T_9 
=0 \), the final hoped-for result being the homology group \(\ker T_8 / \im 
T_9\). A careful implementation of the KB algorithm was available in our 
environment, giving for example the Smith reduction of \(T_8\) in 4 seconds. 
But the same algorithm used for \(T_9\) failed. After a few days of runtime 
without any output, it was obvious a devil was hidden somewhere.

It is well known that the first naive algorithms computing the Smith 
reductions very quickly generate huge intermediate integers with thousands of 
digits, often making it impossible for the calculation to terminate in 
reasonable time.

The Smith reduction consists in using elementary operations on the studied 
matrix, without changing its equivalence class, to cancel the non-diagonal 
entries of the matrix. Kannan and Bachem in their article~\cite{knbc} showed 
how a careful and simple organization of the order of the entries to be 
cancelled gives a polynomial algorithm, avoiding the combinatorial explosion 
of the naive methods.

The Hermite and Smith reductions in~\cite{knbc} are obtained only for square 
invertible matrices, but obvious adaptations extend the scope of the KB 
algorithm to the most general situation, for rectangular matrices of arbitrary 
rank. 

A careful tracing of our implementation of the KB algorithm, detailed in this 
text, revealed in fact the devil was hidden in these ``obvious'' adaptations 
to extend the KB organisation of~\cite{knbc} from the square non-singular case 
to arbitrary matrices, rectangular and arbitrary rank.

Once this point is identified, it is easy (obvious!) to add a small complement 
in the organization of the ``generalized'' KB algorithm to dramatically 
improve our implementation in the general case. For example for our 
matrix~\(T_9\), then the result is obtained in a few minutes. 

Two symmetric Hermite reductions are defined. For a square matrix, let us call 
HNF-1 the column-style reduction giving a lower triangular matrix, and HNF-2 
the row-style reduction giving an upper triangular matrix. In the original KB 
algorithm for the Smith reduction, the HNF-1 reduction was privileged, with a 
sort of minimal potion of HNF-2. It happens the gap in our first erroneous 
generalization of the KB algorithm was fixed thanks to an extra dose of HNF-2.

This relatively complicated mixture of HNF-1 and HNF-2 gives another idea. 
Starting with a rectangular matrix \(M_0\) of arbitrary rank, the HNF-1 
reduction produces a first matrix \(M_1\), lower triangular above a rectangle. 
Now let us simply apply the HNF-2 reduction to \(M_1\), this produces a matrix 
\(M_2\), upper triangular. Experience shows it is in general much more 
``reduced'' than \(M_1\). Why not continue this game? We apply now HNF-1 to 
the last matrix \(M_2\), obtaining \(M_3\), and so on. We prove that this 
process converges toward a diagonal matrix immediately giving the Smith 
reduction. The same proof like in~\cite{knbc} establishes this version of the 
KB algorithm is also polynomial, and experience shows after applying this 
method to numerous examples that it is the fastest one. Furthermore 
programming this version is very simple. Allowing easy extensions to other 
situations, for example for matrices of polynomials.

  This organization of the KB algorithm does not use any \emph{modular} 
technology, it uses only left and right multiplications by unimodular 
matrices, mainly some appropriate B\'ezout matrices, so that if the matrices 
\(u\) and \(v\) satisfying \(s = vdu\) are desired, they can be easily 
determined along this version of the KB algorithm, like in the original KB 
algorithm. Cf in~\cite[Section IV-2]{dtlq} the complementary calculations 
which are necessary if the Smith form has been obtained via modular reduction. 
Along the same lines, see the article~\cite{dmsv} how the research of the 
right modulus can be sophisticated, even in favorable situations, in case of 
small \emph{valence}, which cannot be applied here. See also the comments of 
the author after~\cite[Algorithm 2.4.8]{cohn}.

 This point is important in \emph{constructive} homology, when an 
\emph{explicit} \(\bZ\)-cycle is required for some homology class, the very 
basis of \emph{constructive Algebraic Topology}, see Section~\ref{62253}.

About our ``numerous'' examples, an error would consist in testing the various 
versions of the algorithm with banal random matrices. Such matrices are 
generally too simple with respect to the possible difficulties. The same for 
the matrices coming from elementary contexts in algebraic topology, typically 
the homology groups of simplicial complexes; even when these matrices are 
geant, the Smith reduction of these matrices is easy. Our difficult matrix 
\(T_9\) was the result of a sophisticated work in Algebraic Topology, 
explained in the text, and was not at all \emph{arbitrary}, allowing us to 
identify a severe drawback if the KB algorithm is too lazily extended for the 
rectangular matrices of arbitrary rank.

But how it is possible to generate ``interesting'' difficult matrices with 
respect to the Smith reduction? A section of this text is devoted to this 
question, allowing us to easily generate ``difficult'' matrices and to present 
a statistical study of the results of the various versions of the KB algorithm 
with respect to these matrices. This generation process could be used to 
obtain good benchmarks for other Smith reduction algorithms.\vspace{15pt}

As explained in the abstract, an equivalent algorithm is in fact used by the 
program \emph{Magma}, see{\cite{magm}. Just a few lines are given in this 
program handbook, so that the present text, including detailed explanations, 
could be useful.\vspace{10pt}

\noindent{\large\textbf{Plan:}}\vspace{10pt}

Section~\ref{50401} recalls the key tool of the reduction, known as the 
\emph{B\'ezout} matrices.

Section~\ref{96061} explains the original KB algorithm, defined only for the 
non-singular square matrices.

Section~\ref{93904} gives the obvious complements extending the KB algorithm 
to the general case of a rectangular matrix of arbitrary rank.

Section~\ref{90133} describes the problem of Algebraic Topology producing, via 
our \emph{constructive} methods, boundary matrices giving the homology 
group~\(H_8\) of a relatively complicated topological space.

Section~\ref{57376} explains the observed ``accident'' when we tried to apply 
our extended KB algorithm to the matrix \(T_9\), a matrix \(1995 \times 5796\). 
The reason of this accident is described and a simple solution to overcome it 
is given, a successful one.

Section~\ref{94298} observes the solution so obtained is nothing but a mixture 
a little complicated of the two classical Hermite reductions HNF-1 and 
\mbox{HNF-2}. This gives the idea of a direct iterative combination of HNF-1 
and HNF-2. The algorithm then obtained is quite simple.

The last Section~\ref{91676} relates various tests illustrating that this last 
version of the KB algorithm is the fastest one. Designing interesting 
benchmark matrices is not simple, we explain how to obtain such matrices.

\section{B\'ezout matrices.}\label{50401}

The Hermite and Smith reductions consist in applying the classical Euclid's 
algorithm. If \(a\) and \(b\) are non-null integers, then the so called 
extended Euclidean algorithm returns three integers \(p\), \(q\) and \(r\) 
satisfying \(ap + bq = r\) with \(r\) the GCD of \(a\) and \(b\); we may also 
require \(|p| \leq |b|/r\) and \(|q| < |a|/r\) minimal~\cite[Algorithm 
2.4.5]{cohn}, important to master the size of the future integers. Dividing 
the B\'ezout relation by \(r\) gives the relation \(p(a/r) + q(b/r) = 1\), so 
that the matrix:
\begin{equation}
\begin{pmatrix}
 p & -b/r \\ q & a/r
\end{pmatrix}%
\end{equation}%
is unimodular, producing the equivalence:
\begin{equation}
\begin{pmatrix}
 a & b \\ \ast & \ast
\end{pmatrix}%
 \cong
\begin{pmatrix}
 a & b \\ \ast & \ast
\end{pmatrix}%
\begin{pmatrix}
 p & -b/r \\ q & a/r 
\end{pmatrix}
=
\begin{pmatrix}
 r & 0 \\ \ast & \ast
\end{pmatrix}%
\end{equation}%
and we are happy because the entries \(a\) and \(b\) of the initial matrix are 
replaced in an equivalent matrix by \(r\) and 0, so we have cancelled the 
non-diagonal entry \(b\) and are closer to a diagonal equivalent matrix. In 
the same way:
\begin{equation}
\begin{pmatrix}
 a & \ast & b \\ \ast & \ast & \ast \\ \ast & \ast & \ast
\end{pmatrix}
\cong
\begin{pmatrix}
 a & \ast & b \\ \ast & \ast & \ast \\ \ast & \ast & \ast
\end{pmatrix}
\begin{pmatrix}
 p & 0 & -b/r \\ 0 & 1 & 0 \\ q & 0 & a/r
\end{pmatrix}%
=
\begin{pmatrix}
 r & \ast & 0 \\ \ast & \ast & \ast \\ \ast & \ast & \ast 
\end{pmatrix}%
\end{equation}%
with the same sort of result.

We will denote in general by \(e_{i,j}\) the \underline{e}ntry at position 
\((i,j)\) of the ``current'' matrix.

If \(1 \leq i \neq j \leq m\), we call the \emph{column} B\'ezout matrix \(B = 
CB(n, i, j, p, -b/r, q, a/r)\) the square \(m \times m\) identity matrix 
except the entries \(e_{i,i} = p\), \(e_{i,j} = -b/r\), \(e_{j,i} = q\) and 
\(e_{j,j} = a/r\) where \(a, b, p, q\) and \(r\) are the integers of a 
B\'ezout relation \(ap + bq = r\). For example the \(3 \times 3\) B\'ezout 
matrix above would be denoted by \(CB(3,1,3, p, -b/r,q, a/r)\). Right 
multiplying a matrix having \(a\) on the diagonal and \(b\) on the same 
\emph{row} by the appropriate B\'ezout matrix produces an equivalent matrix 
where the entry~\(b\) is cancelled.

In the same way, with the same B\'ezout relation, the multiplication of 
matrices:
\begin{equation}
\begin{pmatrix}
 p & q \\ -b/r & a/r
\end{pmatrix}%
\begin{pmatrix}
 a & \ast \\ b & \ast
\end{pmatrix}%
= \begin{pmatrix} 
 r & \ast \\ 0 & \ast
\end{pmatrix}%
\end{equation}%
cancels the entry \(b\) on the same \emph{column} as \(a\). So the \emph{row} 
B\'ezout matrix \(RB(n, i, j, p, q, -b/r, a/r)\) defined in the same way as 
\(CB\) can be used for a left multiplication of the matrix \(M\) where 
\(e_{i,i} = a\) and \(e_{j,i} =b\) to produce an equivalent matrix where \(b\) 
on the same \emph{column} as \(a\) is cancelled. 

A particular case is important. If \(a\) divides \(b\), then the B\'ezout 
relation between \(a\) and \(b\) is simply \(1 \cdot a + 0 \cdot b = a\), 
giving the equivalence:
\begin{equation}
\begin{pmatrix}
 a & b \\ \ast & \ast
\end{pmatrix}%
 \cong
\begin{pmatrix}
 a & b \\ \ast & \ast
\end{pmatrix}%
\begin{pmatrix}
 1 & -b/a \\ 0 & 1 
\end{pmatrix}
=
\begin{pmatrix}
 a & 0 \\ \ast & \ast
\end{pmatrix}%
\end{equation}%
So that in this particular case the diagonal entry \(a\) is unchanged. It is 
in fact a column operation which subtracts from the \(b\) column \(b/a\) times 
the~\(a\) column. It is the only case where the diagonal entry is unchanged. 
On the contrary, if \(a\) does not divide \(b\), the diagonal entry \(a\) is 
replaced by the GCD~\(r\) of \(a\) and \(b\) with \(r < a\); the integer \(r\) 
is a \emph{strict} divisor of \(a\).

More generally, we call \(CO(n,i,j, \alpha)\) the column operation consisting 
in subtracting \(\alpha\) times the column \(i\) from the column \(j\), which 
amounts to a right multiplication by an unimodular matrix as explained above. 
This can be used in particular to take account of the Euclidean division 
\(e_{i,j} = e_{i,i} \cdot q + r\) to replace the entry \(e_{i,j}\) by the 
entry \(r\) satisfying \(0 \leq r < e_{i,i}\). The row operation 
\(RO(n,i,j,\alpha)\) is defined in the same way.

These operations can be used also to transform a diagonal matrix into another 
one which satisfies the divisor condition. Consider this sequence of 
equivalences where \(r\) (resp. \(s\)) is the GCD (resp. LCM) of \(a\) and 
\(b\):
\begin{equation}
\begin{pmatrix}
 a & 0 \\ 0 & b
\end{pmatrix}%
\cong
\begin{pmatrix}
 a & b \\ 0 & b
\end{pmatrix}%
\begin{pmatrix}
 p & -b/r \\ q & a/r
\end{pmatrix}%
=
\begin{pmatrix}
 r & 0 \\ bq & ab/r
\end{pmatrix}%
\cong
\begin{pmatrix}
 r & 0 \\ 0 & s
\end{pmatrix}%
\end{equation}%

A row operation has given a \(e_{1,2} = b\), then a column B\'ezout operation 
makes the GCD \(r\) appear, if as usual \(ap + bq = r\); finally \(r\) divides 
\(bq\) and a row operation cancels \(bq\); also \(s = ab/r\) is the LCM of 
\(a\) and \(b\) and is divisible by \(r\). We call this an operation of 
\emph{divisor normalization}. An iteration of such operations allows us, 
without changing the equivalence class, to replace an arbitrary diagonal 
matrix by another one which satisfies the \emph{divisor condition}.

\section{The KB algorithm.}\label{96061}

\subsection{Cancelling the entries above the diagonal.} \label{14210}

The KB algorithm reduces first the given square non-singular matrix \(M_0\) to 
a lower triangular matrix \(M_1\), equivalent to \(M_0\). In particular, the 
determinant of \(M_0\) therefore is the product of the diagonal entries of 
\(M_1\). It is the so-called column-style Hermite reduction, let us call it 
HNF-1.

The tempting way to obtain this reduction consists in using B\'ezout 
operations to cancel the entries above the diagonal in the \emph{most obvious 
order}, described in the case of a \(5 \times 5\) matrix as follows: 
\begin{equation}
\begin{pmatrix}
 \ast & 1 & 2 & 3 & 4\\ \ast & \ast & 5 & 6 & 7 \\
 \ast & \ast & \ast & 8 & 9 \\ \ast & \ast & \ast & \ast & 10 \\
 \ast & \ast & \ast & \ast & \ast
\end{pmatrix}%
\end{equation}%
If an entry to be cancelled is already null, no operation at all for this 
entry, go to the next entry. When starting a row, we must check the diagonal 
entry of this row is non-null and positive. If negative we multiply the column 
by~\(-1\). If null, because the matrix is non-singular, an entry on the same 
row on the right of the diagonal must be non-null. Exchanging two columns, 
which does not change the equivalence class, the null entry of the diagonal is 
replaced by a non-null entry; \emph{with this organization}, the part of this 
new column above the diagonal entry is null, which will no longer be true in 
the next organization.

This is sufficient for student exercises, but with big matrices, combinatorial 
explosions are often generated, needing intermediate entries in the process 
with thousands of digits, leading frequently to a fail of the algorithm, even 
with powerful computers.

The Hermite reduction according to Kannan and Bachem consists in using a 
\emph{different order} for cancelling the entries above the diagonal:
\begin{equation} \label{46491}
\begin{pmatrix}
 \ast & 1 & 2 & 4 & 7 \\ \ast & \ast & 3 & 5 & 8 \\
 \ast & \ast & \ast & 6 & 9 \\ \ast & \ast & \ast & \ast & 10\\
 \ast & \ast & \ast & \ast & \ast 
\end{pmatrix}%
\end{equation}%

So that we cancel successively \(e_{1,2}\), \(e_{1,3}\), \(e_{2,3}\), 
\(e_{1,4}\), \(e_{2,4}\), \(e_{3,4}\) etc. \emph{After} such a B\'ezout 
operation killing the entry \(e_{i,j}\) with \(i < j\), KB uses for every \(1 
\leq k < i\) the Euclidean division \(e_{i,k} = q_{i,k} e_{i,i} + r_{i,k}\) 
and the column operation \(CO(n, i, k, - q_{i,k})\) replaces also \(e_{i,k}\) 
by \(r_{i,k}\) with \(0 \leq r_{i,k} < e_{i,i}\). This process possibly 
decreasing the \(e_{i,k}\)'s for \(k < i\) is repeated every time \(e_{i,i}\) 
is used to cancel an entry \(e_{i,j}\) for \(j > i\); this is useful for the 
lefthand side of the \(i\)-th row, which will be used later for further 
operations.

When the last entry \(e_{n-1, n}\) above the diagonal is cancelled, there 
remains to use the last diagonal entry \(e_{n,n}\) to replace the entries of 
the last row by Euclidean rests of division by \(e_{n,n}\). The new matrix is 
the column-style Hermite reduction of the initial matrix. It is well known 
this Hermite form is unique. Kannan and Bachem proved in~\cite{knbc} this 
algorithm is polynomial. Experience shows it is quite efficient.

\subsection{Canceling the entries below the diagonal.}\label{42039}

It's a little more complicated. We start with a column-style Hermite matrix, 
every entry \(e_{i,j}\) with \(j > i\) is null, all the entries of the 
diagonal are positive, and the entries on the left of the diagonal are 
positive or null, bounded by the corresponding diagonal entry.

Important: the product of the diagonal entries is the absolute value of the 
determinant of the initial matrix, a value which is the same for every 
triangular matrix with positive diagonal entries equivalent to the initial one.

\subsubsection{Phase 1.}

First we get rid of the entries below \(e_{1,1}\)
\begin{equation}
\begin{pmatrix}
 e_{1,1} & 0 & 0 & 0 & 0 \\ 1 & \ast & 0 & 0 & 0\\
 2 & \ast & \ast & 0 & 0\\ 3 & \ast & \ast & \ast & 0 \\
 4 & \ast & \ast & \ast & \ast
\end{pmatrix}%
\end{equation}

The entries \(e_{2,1}\) up to \(e_{n,1}\) are cancelled  with row B\'ezout 
operations based on \(e_{1,1}\). This does cancel these entries with a 
drawback: maybe new non-null entries have appeared above the diagonal:
\begin{equation}
\begin{pmatrix}
 \ast & \ast & \ast & \ast & \ast \\ 0 & \ast & 0 & 0 & 0\\
 0 & \ast & \ast & 0 & 0\\ 0 & \ast & \ast & \ast & 0 \\
 0 & \ast & \ast & \ast & \ast
\end{pmatrix}%
\end{equation}

\subsubsection{Phase 2.}

We then apply again HNF-1 to obtain something like:
\begin{equation}
\begin{pmatrix}
 \ast & 0 & 0 & 0 & 0 \\ \ast & \ast & 0 & 0 & 0\\
 \ast & \ast & \ast & 0 & 0\\ \ast& \ast & \ast & \ast & 0 \\
 \ast & \ast & \ast & \ast & \ast
\end{pmatrix}%
\end{equation}

Looks like a vicious circle? Not at all. If \(e_{1,1}\) does not divide one of 
the entries of the first column, the row B\'ezout operation replaces 
\(e_{1,1}\) by a strict divisor. If on the contrary \(e_{1,1}\) divides the 
entry \(e_{j,1}\), then the B\'ezout operation directly cancels this 
\(e_{j,1}\), nothing else.

\subsubsection{Iteration.}

So that we iterate this process on the column 1; in one iteration, during the 
phase 1, either \(e_{1,1}\) is replaced by a strict divisor, either it is 
unchanged in which case this means it divides \emph{all} the non-null entries 
of the first column, which entries are directly cancelled, without new 
non-null entries on the first row. The possible successive values of 
\(e_{1,1}\) are bounded from below by~1, so that when the value of \(e_{1,1}\) 
becomes fixed, all the other entries of the column 1 and the row 1 are null. 
We have obtained a matrix equivalent to the original one which looks like:
\begin{equation}
\begin{pmatrix}
 \ast & 0 & 0 & 0 & 0 \\ 0 & \ast & 0 & 0 & 0\\
 0 & \ast & \ast & 0 & 0\\ 0 & \ast & \ast & \ast & 0 \\
 0 & \ast & \ast & \ast & \ast
\end{pmatrix}%
\end{equation}

The same work can then be done based on \(e_{2,2}\), then \(e_{3,3}\) and so 
on. Finally, we obtain a diagonal matrix equivalent to the initial one, and a 
divisor normalization can be applied to obtain the final matrix where the 
divisor condition is satisfied.

The original KB algorithm used a simple intermediate complement to directly 
obtain a diagonal with the divisor condition satisfied. This is a detail, 
without any devil, and we prefer this version which better prepares us to our 
final version.

\section{Obvious extensions.}\label{93904}

We add now a few obvious adaptations to extend the KB algorithm to the general 
case of a rectangular matrix \(M: \bZ^m \rightarrow \bZ^n\) of arbitrary rank. 
The integer \(m\) (resp. \(n\)) is the column (resp. row) number of our 
matrix. We call \(k\) the \emph{rank} of \(M\), an integer \(k \leq 
\min(m,n)\).

\subsection{First adaptation.}

When we treat the column \(i\) in the first Hermite step, we cancel the 
entries \(e_{1,i}\) to \(e_{i-1,i}\) above \(e_{i,i}\), using the diagonal 
entries \(e_{1,1}\) to \(e_{i-1,i-1}\). It is then possible \(e_{i,i}\) is 
null, a problem for the rest of the process. When the matrix is non-singular, 
a non-null entry \(e_{i,j}\) for \(j > i\) is certainly present, in which case 
an exchange of columns solves the problem.

In the general case, the organization is a little different. We look 
\emph{first} for a non-null \(e_{j,i}\) for \(j > i\) \emph{below} 
\(e_{i,i}\). If such a non-null \(e_{j,i}\) is found, we exchange the 
\emph{rows} \(i\) and \(j\) to install this \(e_{j,i}\) in position \((i,i)\). 
This is forbidden in the \emph{column-style} Hermite reduction: exchanging two 
rows amounts to \emph{left} multiplying the current matrix by a permutation 
matrix. In other words, the matrix we will obtain in this way maybe is not the 
unique column-style  Hermite reduction of the original matrix. But we are 
interested in fact by the Smith reduction, so that this is not a real problem. 
\emph{Even if this happens we continue to call HNF-1 the process so defined}.

If such a non-null \(e_{j,i}\) is found, now installed in position \((i,i)\), 
we continue as before. If no such \(e_{j,i}\) is found, this means the column 
\(i\) is now entirely null. \emph{Then} we look for a non-null \emph{column} 
\(j\) with \(j > i\) to the right of \(e_{i,i} = 0\) If such a column is 
found, we exchange the columns \(i\) and \(j\) and retreat this column as 
before  restarting from the position \(e_{1,i}\).

If no non-null column to the right of \(e_{i,i}\) is found this means all the 
columns \(i\) to \(m\) are now null. This proves the rank \(k\) of the initial 
matrix is in fact \(k = i-1\) and we stop there the HNF-1 process. We have so 
obtained a lower triangular matrix of rank \(k\) above an arbitrary rectangle 
\(n-k\) rows and \(k\) columns. The rows of this rectangle are 
\(\bQ\)-generated by the \(k\) rows of the triangular matrix, but not in 
general \(\bZ\)-generated. The format of this matrix is:{\footnotesize
\begin{equation}\label{31645}
\begin{pmatrix}
 e_{1,1} & 0       & 0       & \cdots & 0       & 0      & \cdots & 0 \\
 \ast    & e_{2,2} & 0       & \cdots & 0       & 0      & \cdots & 0 \\
 \ast    & \ast    & e_{3,3} & \cdots & 0       & 0      & \cdots & 0 \\
 \vdots  & \vdots  & \vdots  & \ddots & \vdots  & \vdots & \ddots & \vdots \\
 \ast    & \ast    & \ast    & \cdots & e_{k,k} & 0      & \cdots & 0 \\
 \ast    & \ast    & \ast    & \cdots & \ast    & 0      & \cdots & 0 \\
 \vdots  & \vdots  & \vdots  & \ddots & \vdots  & \vdots & \ddots & 0 \\
 \ast    & \ast    & \ast    & \cdots & \ast    & 0      & \cdots & 0
\end{pmatrix}%
\end{equation}}%
where the stars are some integers.

The product \(G' = e_{1,1} \cdots e_{k,k} \) of the diagonal entries of the 
almost Hermite matrix obtained is a multiple of \(G\), this \(G\) being the 
GCD of the determinants of all the \(k \times k\) minors of the initial 
matrix. This \(G\), well defined for the initial matrix, the same for all the 
intermediate matrices up to the Smith form, constant, will play an essential 
role, keep it in mind. You can see a small example at the end of 
Section~\ref{16739}.

\subsection{Second adaptation.}

If less columns than rows, if \(m < n\), it is possible \(k = m\), in which 
case the Hermite diagonal finishes in position \((k,k)\), nothing more is to 
be done for the HNF-1 step.

If more columns than rows, if \(m > n\) and if \(k = n\), when the Hermite 
process \`a la KB has obtained \(e_{k,k}\), it remains the arbitrary columns 
\(k+1\) to \(m\) to process. This is done in the same way as before, using the 
non-null diagonal entries already obtained to cancel all the entries 
\(e_{i,j}\) for \(j > k\) of these columns. In general this decreases the 
diagonal entries. We finally have a triangular matrix which is the correct 
column-style Hermite reduction of the original matrix. In this case, \(G' = 
G\).

If \(k < n < m\), the most frequent case in ``hard'' algebraic topology, when 
we have obtained the first version of \(e_{k,k}\), the story continues as 
follows. We process the column \(k+1\), cancelling the entries \(e_{1,k+1}\) 
to \(e_{k,k+1}\), using the available diagonal entries \(e_{1,1}\) to 
\(e_{k,k}\). These diagonal entries often are replaced by strict divisors. 
Then surprise, the column \(k+1\) is become entirely null, otherwise the rank 
would be \(> k\). So that we look for a non-null column at position \(j > 
k+1\); when it is found it is exchanged with the column \(k+1\) now null, and 
this new column \(k+1\) is processed in the same way, same surprise, and so on 
finishing with a matrix as in~(\ref{31645}).

\subsection{End of the computation.}\label{67573}

When the non-null matrix finally obtained is triangular, the same process as 
the one described in~\cite{knbc} can be used, nothing is changed; see 
Section~\ref{42039}.

When we also get a non-null rectangle below the triangle, nothing is changed 
for the last step either. The columns are only  more or less higher, but 
because the row-style Hermite reduction is used column by column, the same 
process does give the hoped-for Smith form. 

\subsection{Computing a constructive homology group.} \label{62253}

Let a homology group \(H\) be obtained via a quotient \(H = \ker d' / \im d\). 
In \emph{constructive} homological algebra, it is not only necessary to know 
that for example \(H = \bZ/12\), but if \(g \in \bZ/12\) is a generator, what 
element of \(\ker d'\) could represent it? This problem is solved through the 
auxiliary matrices \(u', v', u, v\) which complement the Smith reductions 
\(s'\) and \(s\).

More precisely, let us consider this diagram:
\begin{equation}
\begin{tikzpicture} [xscale = 2, yscale = 1.6, baseline = (s)]
 \node (03) at (0,3) {\(\bZ^m\)} ;
 \node (02) at (0,2) {\(\bZ^m\)} ;
 \node (13) at (1,3) {\(\bZ^n\)} ;
 \node (12) at (1,2) {\(\bZ^n\)} ;
 \node (11) at (1,1.4) {\(\bZ^k\)} ;
 \node (10) at (1,0.4) {\(\bZ^k\)} ;
 \node (23) at (2,3) {\(\bZ^p\)} ;
 \node (21) at (2,1.4) {\(\bZ^p\)} ;
 \node (s) [rotate = 90] at (0.95,1.7) {\(\subset\)} ;
 \begin{scope} [->, font = \scriptsize]
 \draw (13) -- node [above] {\(d'\)} (03) ;
 \draw (23) -- node [above] {\(d\)} (13) ;
 \draw (03) -- node [left] {\(v'\)} (02) ;
 \draw (12) -- node [above] {\(s'\)} (02) ;
 \draw (12) -- node [left] {\(u'\)} (13) ;
 \draw (23) -- node [inner sep = 2pt, rotate = 53.3, above] {\(u'^{-1}d\)} (11) ;
 \draw (11) -- node [left] {\(v\)} (10) ;
 \draw (21) -- node [right] {\(u\)} (23) ;
 \draw (21) -- node [inner sep = 2pt, rotate = 40, above] {\(s\)} (10) ; 
 \end{scope}%
\end{tikzpicture}%
\end{equation}%
with \(d'd = 0\). The Smith reduction of \(d'\) produces \(s'\), \(u'\) and 
\(v'\) as in the diagram. The matrix \(s'\) is diagonal so that the kernel 
\(\bZ^k\) of \(s'\) is generated by the last \(k\) factors of \(\bZ^n\). Now 
\(d'd=0\) implies \(s' u'^{-1} d = 0\) and the image of \(u'^{-1}d\) is in 
\(\bZ^k\). The Smith reduction of \(u'^{-1}d\) produces \(s\), \(u\) and 
\(v\). The generators of the homology group \(H = \ker d' / \im d\) correspond 
to the basis vectors of the lower \(\bZ^k\) whose corresponding diagonal entry 
in \(s\) is not 1. It remains to apply \(u'v^{-1}\) to these basis vectors to 
produce the cycles in \(\ker d'\) which represent the generators of \(H\).

In other words, if the matrices \(u', v', u, v\) are not available, the cycles 
representing the homology classes are not reachable and the methods of 
\emph{constructive} homology cannot be applied.

\section{A problem of Algebraic Topology.}\label{90133}

This section assumes a minimal knowledge in Algebraic Topology and can be 
skipped by the readers interested only by the algebraic problem of the Smith 
reduction.

The standard methods of Algebraic Topology, mainly the exact and spectral 
sequences, are \emph{not} algorithms producing for example the desired 
homology or homotopy groups. They \emph{sometimes} succeed in determining 
\emph{some} groups but never have a general scope.

This problem led the author and the colleagues of his research group to design 
which is called now the \emph{Constructive Algebraic Topology}, 
see~\cite{rbsr06}.

Relatively simple examples have been used to illustrate the power of these 
methods. Let us start with the infinite real projective space 
\(P^\infty(\bR)\); it's the inductive limit of the \(P^n(\bR)\)'s, so the 
inclusion relation \(P^3(\bR) \subset P^\infty\bR)\) can be used to define the 
quotient space \(P_4 := P^\infty(\bR) / P^3(\bR)\). It is elementary to proof 
\(P_4\) is 3-connected and \(\pi_4(P_4) = \bZ\). Therefore the first non-null 
homotopy group of the loop space \(\Omega(P_4)\) is \(\pi_3 = \bZ\).

This implies attaching a 4-cell \(e^4\) to \(\Omega(P_4)\) by a map \(S^3 
\rightarrow \Omega(P_4)\) of degree 4 makes sense, producing the space 
\(\Omega(P_4) \cup_4 e^4\). Now \(\pi_3(\Omega(P_4)~\cup_4 e^4) = \bZ/4\). The 
loop space of the last space is simply connected and 
\(\pi_2(\Omega(\Omega(P_4) \cup_4 e^4)) = \bZ/4\). Attaching a 3-cell \(e^3\) 
to the last space by a map \(S^2 \rightarrow \Omega(\Omega(P_4) \cup_4 e^4)\) 
of degree 2 makes sense, and we take again the loop space of the last space, 
obtaining finally the space:
\begin{equation}
 X = \Omega(\Omega(\Omega(P^\infty(\bR) / P^3(\bR)) \cup_4 e^4) \cup_2 e^3)
\end{equation}%

Question: what about the homology groups of \(X\)?

This space \(X\) has been chosen because it cumulates well known 
``difficulties'' of standard algebraic topology. The homology groups of a loop 
space are generally believed be reachable through the Eilenberg-Moore spectral 
sequence. But this spectral sequence is \emph{not}, in the standard context, 
an algorithm.

If the space \(S\) is a simplicial set of \emph{finite type}, Franck Adams' 
Cobar construction~\cite{adms} \emph{is}, in modern language, an algorithm 
computing the homology groups of the \emph{first} loop space. But for example 
\(\Omega(P_4\)) is not of finite type, so that the Cobar construction does not 
give the homology groups of the second loop space \(\Omega(\Omega(P_4)) =: 
\Omega^2(P^4)\).

Twenty-four years later, Hans Baues' AMS Memoirs~\cite{baus2} is entirely 
devoted to an algorithm which computes the homology groups of the second loop 
space \(\Omega^2(S)\) of a simplicial set of finite type. The second loop 
space only. This is the reason why in our example \(X\), we have chosen to 
apply three times the loop space functor.

In fact, the general problem of the homology groups of \(\Omega^n(S)\), for 
\(S\) of finite type and \(n\)-connected, was solved in Julio Rubio's 
thesis~\cite{rubi2}, thanks to the methods of Constructive Algebraic Topology.

Let \(S\) be an arbitrary space and let us assume we know the homology groups 
of the first loop space \(\Omega S\). If we attach a cell \(e^n\) to \(S\) by 
a map \(f: S^{n-1} \rightarrow S\), what about the homology groups of 
\(\Omega(S \cup_f e^n)\)? A spectral sequence is available for example 
in~\cite[Section III.2]{baus3}, but this is not an algorithm computing these 
homology groups. On the contrary, if the \emph{effective} homology of \(S\) is 
known, which makes sense even if \(S\) is not of finite type, then the methods 
of Constructive Algebraic Topology give an \emph{algorithm} computing the 
homology groups of \(\Omega(S \cup_f e^n)\). In particular this covers the 
attachments in the definition of our \(X\).

Also, in the particular case the space \(S\) is an \(n\)-th suspension, then a 
(true) algorithm is known~\cite{mlgr} giving the homology groups of \(\Omega^n 
S\). This is the reason we have chosen \(P_4 = P^\infty(\bR) / P^3(\bR)\) as 
the initial space: this space is in a sense the simplest example of a 
3-connected space which is not a suspension.

The methods of Constructive Algebraic Topology, are not only theoretical 
algorithms with a large scope, but they are concretely implemented in the 
program Kenzo~\cite{drss}. And to illustrate the power of these methods and of 
this program, we tried to compute the homology groups of \(X\), our a little 
contorted space.

In this case the Kenzo program worked during about one week on a relatively 
powerful computer, briefly described in the Appendix, to produce the two 
matrices \(T_8\) and \(T_9\) mentioned in the introduction. These matrices are 
such the desired homology group \(H_8 X\) is the quotient group \(\ker T_8 / 
\im T_9\), which group is deduced from the Smith reduction of \(T_8\) and 
\(T_9\). The final result is:
\begin{equation}
 H_8 X = (\bZ/2)^{253} + (\bZ/4)^9 + \bZ/8 + \bZ^5
\end{equation}%
But what about the Smith reduction of these matrices?

\section{An accident, a solution.}\label{57376}

We used the obvious KB extension described in Section~\ref{93904} for years 
for many matrices, sometimes relatively large, without any problem. The first 
accident happens for the matrix \(T_9\), a matrix \(1995 \times 5796\) with 
24374 non-null entries in \([-124\ldots 132] \subset \bZ\), about 97.9\% of 
the entries are null.

Tracing carefully the work of our obvious extension of the KB algorithm, we 
discovered a terrible growth of the entries of the inferior rectangle. After 
the HNF-1 step, we have a matrix with a lower triangle of height and basis \(k 
= 1481\), and below a rectangle \(514 \times 1481\). When the phase 2 is in 
work, the program processes the first 500 columns in one minute and a half, 
but the \emph{average} number of \emph{digits} of the entries under the 
diagonal is already~276. After a few days, the column 682 was processed, which 
took about 15 hours to be processed, for one column only. The \emph{average} 
number of digits of the entries under the diagonal is then 482222. And 189217 
entries are yet to be cancelled, 799 columns are remaining to be processed. It 
is clear our obvious extension of the KB algorithm meets a problem.

The explanation of this accident is simple. When we cancel, see 
Section~\ref{42039}, the entries of the column \(i\) that are below the 
diagonal entry \(e_{i,i}\), we apply the HNF-2 reduction only to this column, 
and then we apply the HNF-1 reduction to the new matrix. In particular, the 
diagonal entries \(e_{1,1}\) to \(e_{k,k}\) are used to decrease the other 
entries of the same \emph{rows}, but this does not decrease the entries of the 
lower rectangle. Which entries continue to grow without any limit, giving the 
standard combinatorial explosion which easily happens if no precautions are 
taken to master such a growth.

The solution is very simple. These diagonal entries \(e_{1,1}\) to \(e_{k,k}\) 
are used to decrease the entries of the \emph{rows} 1 to \(k\) when the HNF-1 
reduction is applied; after this action, every entry of such a row is bounded 
by the corresponding diagonal term. Unfortunately, this is without any effect 
on the rows \(k+1\) to~\(n\). Why not use these diagonal entries to decrease 
in the same way the entries of the corresponding \emph{columns}, using 
elementary \emph{row operations}? After such an action every entry of the 
matrix is bounded by the corresponding diagonal entry of the same column. In 
the whole process, the product of the diagonal entries is a divisor of \(G'\), 
so that we are so sure to prevent the indefinite growth of the entries of the 
lower rectangle.

It happens adding this small complement after the processing of every column 
in the second part of the KB algorithm is enough to make reasonably efficient 
the extended KB algorithm. This done, this updated algorithm gives in 
\(\sim12\) minutes the Smith reduction of our matrix \(T_9\):
\begin{equation}
 ((1218 \ast 1) (253 \ast 2) (9 \ast 4) (1 \ast 8))
\end{equation}%
meaning the diagonal is made of 1218 entries 1, 253 entries 2, and so on.

But it is still possible to do better.

\section{A simpler solution.}\label{94298}

The last reasonably efficient version of the extended KB algorithm consisted 
in adding a small dose of HNF-2: the HNF-2 reduction \`a la KB would 
systematically use such simple row operations to decrease the entries on the 
same column below a diagonal entry \(e_{i,i}\).

This gives another idea. Why not use simply alternately the HNF-1 and the 
HNF-2 Smith reductions? This idea was in fact already present in the initial 
KB algorithm when processing a column in the second part of the algorithm.

In the general case, starting from a matrix \(M_0\) the HNF-1 reduction gives 
a lower triangle above a rectangle in a matrix \(M_1\), the rank \(k\) is now 
known. If we apply HNF-2 to this matrix \(M_1\), we will get only an upper 
triangle in a matrix \(M_2\) where the product of the diagonal entries is then 
exactly the GCD \(G\) of all the \(k \times k\) minors of the initial matrix. 
We then apply HNF-1 to \(M_2\), obtaining the matrix \(M_3\), now lower 
triangular, and so on.

\begin{thr}
 After a finite number \(\nu\) of steps, the matrix \(M_\nu\) obtained is diagonal.
\end{thr}%

\prof All the obtained matrices \(M_\mu\) for \(\mu \geq 2\) are triangular 
and the product of the diagonal entries is constant equal to \(G\).

Let us assume for example we apply HNF-1 to an upper triangular matrix. All 
the diagonal terms are positive. We must apply a number of column B\'ezout 
operations to cancel the entries above the diagonal, processing the columns 
from left to right as explained in Section~\ref{14210}, see the 
matrix~(\ref{46491}).

But our initial matrix is upper triangular, so that before using the diagonal 
entry \(e_{i,i}\) to cancel the entry \(e_{i,j}\) with \(j > i\), the 
symmetric entry \(e_{j,i}\) below the diagonal is \emph{null}, and the main 
part of the B\'ezout operation at the positions \((i,i)\), \((i,j)\), 
\((j,i)\) and \((j,j)\) is particular:
\begin{equation}
 \begin{pmatrix}
  a & b \\ c=0 & d
 \end{pmatrix}
 \begin{pmatrix}
  p & -b/r \\ q & a/r
 \end{pmatrix}%
 =
 \begin{pmatrix}
  r & 0 \\ dq & ad/r
 \end{pmatrix}%
\end{equation}%
where as usual \(ap+bq =r\) is the B\'ezout relation between \(a\) and \(b\). 
The product of both diagonal terms that are concerned is now \(ad\) unchanged. 
So that in this process, the product of the diagonal entries remains constant; 
let us also remark that, when the column \(i\) has been entirely processed, 
the state of the matrix is block  \(2 \times 2\) where the left upper block is 
lower triangular, the right lower block is upper triangular, the left lower 
block is null and it is not amazing the product of the diagonal terms is 
unchanged; for example when the columns 2 and 3 have been processed for an 
upper triangular \(5 \times 5\) matrix, the situation is as follows:
\begin{equation}
\begin{pmatrix}
\begin{array}{ccc|cc}
 \ast & 0 & 0 & \ast & \ast \\
 \ast & \underline{\ast} & 0 & \underline{\ast} & \ast \\
 \ast & \ast & \ast & \ast & \ast \\
 \hline
 0 & \underline{0} & 0 & \underline{\ast} & \ast \\
 0 & 0 & 0 & 0 & \ast
\end{array}%

\end{pmatrix}%
\end{equation}%

If \(e_{i,i}\) divides \(e_{i,j}\), then \(e_{i,i}\), \(e_{j,i} = 0\) and 
\(e_{j,j}\) are unchanged, \(e_{i,j}\) is cancelled; in the matrix above, the 
entries corresponding to \(i = 2\) and \(j = 4\) are underlined. In 
particular, if \(e_{i,j} = 0\), no operation at all is applied for this pair 
\((i,j)\).

If the B\'ezout relation is non-trivial, \(e_{i,i}\) is replaced by a 
\emph{strict} divisor, the missing factor being moved \emph{below} on the 
diagonal, but the product \(G\) of the diagonal entries remains constant.

This game with the divisors of the diagonal entries must stop, and it stops 
only if \emph{all} the diagonal entries divide \emph{all} the corresponding 
other entries, so that the result of this Hermite reduction is diagonal.\qed

When a diagonal matrix is so obtained, a \emph{divisor normalization} as 
explained at the end of Section~\ref{50401} gives the canonical Smith 
reduction.

Experience shows the computing times of the successive HNF invocations quickly 
tend to 0. For example for the matrix \(T_9\) which was our initial example, 3 
HNF invocations are necessary with the successive computing times: 260 - 5 - 0 
seconds. The last 0 means the last HNF was entirely executed in less than 1 
second.

For the matrix \boxtt{Test-15000-1} of the next section, the most difficult of 
our examples, seven HNF invocations are necessary, with the respective 
computing times: 1h31m25s - 14m20s -14m09s - 14s - 1s - 0s - 0s.

\section{Benchmarks}\label{91676}

We intend to compare the three versions of the KB algorithm now available:
\begin{fitem}
\item
KB1: the original KB algorithm of~\cite{knbc} naively extended to rectangular
matrices of arbitrary rank.
\item
KB2: the same algorithm where we use extra row operations to prevent the 
entries of the lower rectangle from indefinite growth.
\item
KB3: the final version where we alternately use the Hermite reductions HNF-1 
and HNF-2 up to obtaining a diagonal matrix.
\end{fitem}%

\subsection{Generating test matrices.} \label{16739}

It was already explained the lazy solution consisting in generating 
rectangular matrices with random entries is erroneous: the Smith reduction is 
then made of 1's with the exception of one or two entries at the end of the 
diagonal. On the contrary, the Smith reduction of our matrix \(T_9\) had 263 
entries \(> 1\) and it's certainly why this matrix raised unexpected 
difficulties. But how to obtain matrices which in a sense are highly Smith 
non-trivial?

There is a simple method, which in particular easily generates examples of 
matrices with the same accident when using the KB1 algorithm.

This method consists in \emph{starting} with a diagonal matrix made of 
arbitrary positive integers, for example random positive integers in an 
interval \([1 \ldots n]\) where \(n\) can be chosen. The rank is predefined as 
the number of non-null entries on the diagonal. Then we remember which is 
usually called the \emph{elementary} operations which can be used on a matrix 
without changing its equivalence class, therefore without changing its Smith 
reduction:
\begin{fitem}
\item
Multiply a column or a row by -1.
\item
Swap two columns or two rows.
\item
Add to some column (resp. row) the product of another column (resp. row) by 
some integer.
\end{fitem}%

The \emph{elementary step} of our generation process is made of five substeps:
\begin{fenum}
\item
Choose two different random columns of our matrix and swap them.
\item
Choose a random column and a random row and multiply these column and row by 
-1.
\item
Choose two different random rows and some random integer \(\alpha \in [-a 
\ldots a]\); then add to the second row \(\alpha\) times the first row.
\item
Choose two different random rows of our matrix and swap them.
\item
Choose two different random columns and some random integer \(\alpha \in [-a 
\ldots a]\); then add to the second column \(\alpha\) times the first column.
\end{fenum}%

A generator of random integers being available, the number \(a\) being given, 
this defines an elementary step of transformation of our matrix. This process 
can be repeated an arbitrary number of times.

A toy example: starting from a \(4 \times 5\) matrix where the diagonal is the 
Smith form \((1\ 3\ 9\ 0)\), applying 10 times our elementary step with \(a = 
10\), using the generator of random integers of our Lisp system, we obtain 
this matrix:{\footnotesize
\begin{equation}
\begin{pmatrix}
 1 & 0 & 0 & 0 & 0 \\
 0 & 3 & 0 & 0 & 0 \\
 0 & 0 & 9 & 0 & 0 \\
 0 & 0 & 0 & 0 & 0
\end{pmatrix}%
\text{\Large\(\longmapsto\)}
\begin{pmatrix}
 37584 & 4383 & 29997 & -54 & 11688 \\
 308 & 36 & 250 & 0 & 96 \\ 
 -40316 & -4707 & -33907 & -153 & -12552 \\ 
 5626 & 657 & 4778 & 27 & 1752
\end{pmatrix}%
\end{equation}}%

As an example, applying the HNF-1 and then the HNF-2 operator to the last 
matrix we obtain successively:
\begin{equation}
\text{\Large\(\stackrel{\text{\footnotesize HNF-1}}{\longmapsto}\)}
\begin{pmatrix}
 3   & 0  & 0    & 0 & 0 \\
 0   & 2  & 0    & 0 & 0 \\
 180 & 49 & 3087 & 0 & 0 \\
 -30 & -8 & -513 & 0 & 0
\end{pmatrix}%
\text{\Large\(\stackrel{\text{\footnotesize HNF-2}}{\longmapsto}\)}
\begin{pmatrix}
 3 & 0 & 0 & 0 & 0 \\
 0 & 1 & 0 & 0 & 0 \\
 0 & 0 & 9 & 0 & 0 \\
 0 & 0 & 0 & 0 & 0
\end{pmatrix}%
\end{equation}%
and it remains to permute the diagonal terms to obtain the canonical Smith 
reduction.

In this example, the product \(G' = e_{1,1}e_{2,2}e_{3,3} = 3 \times 2 \times 
3087 = 686 \times 27 = 686 \times G\) where \(G = 27\) is the GCD of the 
determinants of the 3-minors of the initial matrix, and therefore the product 
of all the divisors of the Smith reduction.

\subsection{Experiments.} \label{96357}

We now work as follows. A list of numbers is given for every experiment. For 
example the first list is \((10 \ 100 \ 300 \ 80 \ 20 \ 300 \ 10)\) with the 
respective meanings:
\begin{fitem}
\item
10: The experiment is repeated 10 times with 10 different matrices as 
described now.
\item
100: Number of rows.
\item
300: Number of columns.
\item
80: Rank.
\item
20: The initial diagonal entries are random integers in \([1 \ldots 20]\).
\item
300: The number of elementary steps that are run where\ldots
\item
10: \ldots the coefficient \(\alpha\) for the row and column operations is a 
random integer \(\alpha \in [-10 \ldots 10]\).
\end{fitem}%

For each experiment, we give the average absolute value of the non-null 
entries of the \emph{first} matrix of the experiment, and also the percentage 
of null entries. We then give the Smith reduction that has been computed for 
\emph{this} matrix. For each version of the KB algorithm, we give the total 
runtime, and for the KB3 version, we give also the number of HNF reductions 
which were necessary.

The appendix gives the necessary references to reach these matrices and the 
corresponding execution listings.

\subsubsection{\((10 \ 100 \ 300 \ 80 \ 20 \ 300 \ 10)\)} \label{62730}

For the first matrix, about 92\% of entries are null and the average absolute 
value of the non-null entries is \(1.91 \times 10^5\). The Smith reduction of 
the first matrix is:
\begin{multline}
((41 \ast 1) (12 \ast 2) (8 \ast 6) (10 \ast 60) (2 \ast 180) (1 \ast 2520) (1 
\ast 42840) 
\\
  (1 \ast 556920) (1 \ast 10581480) (2 \ast 116396280) (1 \ast 232792560))
\end{multline}%
\begin{fitem}
\item
KB1: total runtime = \(2.94\) seconds.
\item
KB2: total runtime = \(2.36\) seconds. 
\item
KB3: total runtime = \(2.02\) seconds. 4 HNF reductions have been necessary 
eight times and 5 HNF two times.
\end{fitem}%

\subsubsection{\((10 \ 150 \ 500 \ 120 \ 10 \ 500 \ 10)\)}

For the first matrix, about 92\% of entries are null and the average absolute 
value of the non-null entries \(5.42 \times 10^6\). The Smith reduction of the 
first matrix is:
\begin{multline}
((57 * 1) (32 * 2) (4 * 6) (2 * 12) (13 * 60) (1 * 180) (11 * 2520))
\end{multline}%
\begin{fitem}
\item
KB1: total runtime = \(23.4\) seconds.
\item
KB2: total runtime = \(10.6\) seconds. 
\item
KB3: total runtime = \(9.8\) seconds. 4 HNF reductions have been necessary 
eight times and 5 two times.
\end{fitem}%

\subsubsection{\((10 \ 500 \ 1500 \ 400 \ 10 \ 1000 \ 10)\)}

For the first matrix, about 99.194\% of entries are null and the average 
absolute value of the non-null entries is \(1.64 \times 10^4\). The Smith 
reduction of the first matrix is:
\begin{multline}
((207 \ast 1) (78 \ast 2) (31 \ast 6) (3 \ast 30) \\ (30 \ast 60) (14 \ast 
180) (6 \ast 360)
  (31 \ast 2520))
\end{multline}%
\begin{fitem}
\item
KB1: total runtime = ??? After two days, it was clear the program failed in a 
reasonable time. The KB1 version is no longer used for the next experiments.
\item
KB2: total runtime = \(106\) seconds. 
\item
KB3: total runtime = \(76\) seconds. 4 HNF reductions have been necessary one 
time, 5 five times and 6 four times.
\end{fitem}%

\subsubsection{\((2 \ 2000 \ 6000 \ 1600 \ 10 \ 1000 \ 10)\)}

For the first matrix, about 99.9743\% of entries are null and the average 
absolute value of the non-null entries is 68. The Smith reduction of the first 
matrix is:
\begin{multline}
((796 \ast 1) (334 \ast 2) (151 \ast 6) (5 \ast 12) \\ (157 \ast 60) (3 \ast 
120) (8 \ast 360)
  (146 \ast 2520))
\end{multline}%
\begin{fitem}
\item
KB2: total runtime = \(133\) seconds. 
\item
KB3: total runtime = \(55.1\) seconds. 4 HNF reductions have been necessary 
one time, and 5 another time.
\end{fitem}%

\subsubsection{\((2 \ 5000 \ 15000 \ 4000 \ 20 \ 8000 \ 10)\)}  \label{44558}

For the first matrix, about 99.956\% of the entries are null and the average 
absolute value of the non-null entries is 55160. 32841 entries are non-null. 
The Smith reduction of this first matrix is:
\begin{multline}
((2009 * 1) (803 * 2) (187 * 6) (181 * 12) (418 * 60) (6 * 180) \\ (186 * 2520)
  (7 * 42840) (5 * 471240) (7 * 17907120) (191 * 232792560))
\end{multline}%
The computing times:
\begin{fitem}
\item
KB2: total runtime = 22 hours.  
\item
KB3: total runtime = 4.4 hours. The first matrix needed six HNF invocations, 
the second one, \boxtt{Test-15000-1}, needed as already explained seven HNF 
steps.
\end{fitem}%

So that KB3 is in this case about 5 times faster than KB2.

Comparing the examples~\ref{62730} to~\ref{44558} makes clear that more 
difficult is the tested matrix, better is the KB3 algorithm with respect to 
KB2. The same kernel functions, HNF reductions, B\'ezout operations and so on, 
have been used in both cases in the same environment for KB2 and KB3 tests, so 
that the difference is only due to the different general organizations.

\subsubsection{T9.}

  We close our ``experiments'' by the current status of our programs with respect 
to the matrix \(T_9\) at the origin of this work. The density of null entries 
is 99.6537\%; the average absolute-value of the non-null terms is simply 
\(3.0275\).

  This matrix is Smith-reduced by KB2 in about 7 minutes and by KB3 in 4.5 minutes. 
In the last case, 3 HNF reductions are enough.

\section{Appendices.}

\subsection{Our machine.}

  The technical data of the computer used for these experiments:
\begin{fitem}
\item
Dell PowerEdge Server R740.
\item
Two Intel processors Xeon Gold 6242R, 3.1GHz, 35.75M Cache, 10.40GT/s, 2UPI, 
Turbo, HT-20C/40T; 40 cores.
\item
RAM = 512Go.
\end{fitem}%

It's an opportunity to thank the engineers of the Computer Center of the 
Fourier Institute, Didier Depoisier and Patrick Sourice, for their patient, 
constant and friendly help.

\subsection{Matrices and listings.}

All the matrices used for experiments in Section~\ref{96357} are available on 
our website~\cite{srgr04}. The \emph{complete} execution listings are also 
included.

\begin{center}
-o-o-o-o-o-o-
\end{center}%

\end{document}